\documentstyle[11pt]{article}
\newcommand{\dl}{\delta^{(3)}({\bf r})}

\newcommand{\BE}{\begin{equation}}
\newcommand{\EE}{\end{equation}}
\newcommand{\BA}{\begin{eqnarray}}
\newcommand{\EA}{\end{eqnarray}}

\begin{document}
\begin{titlepage}

\vspace*{1mm}
\begin{center}

            {\LARGE{\bf Spontaneous symmetry breaking and \\
                        the $p \to 0$ limit }}

\vspace*{14mm}
{\Large  M. Consoli }
\vspace*{4mm}\\
{\large
Istituto Nazionale di Fisica Nucleare, Sezione di Catania \\
Corso Italia 57, 95129 Catania, Italy}
\end{center}
\begin{center}
{\bf Abstract}
\end{center}

We point out a basic ambiguity in the $p \to 0$ limit of the
connected propagator in a spontaneously  broken phase. 
This may represent an indication that the conventional singlet 
Higgs boson, rather than being a purely massive field,
might have a gap-less branch. This
would dominate the energy spectrum for ${\bf{p}} \to 0$ and give rise
to a very weak, long-range force. The natural interpretation is in terms
of density fluctuations of the `Higgs condensate': in the
region of very long wavelengths, infinitely 
larger than the Fermi scale, it cannot be treated as a purely classical 
c-number field. 
\vskip 25 pt
PACS: 11.30.Qc, 14.80.Bn~~~~~~~~~~~~~~~
\vskip 200 pt
(To be published in Physical Review D.)
\vskip 35 pt
\end{titlepage}

\section{Introduction}

The ground state of
spontaneously broken theories is frequently denoted 
as the `Higgs condensate'. In this view, the name 
itself (as for the closely related gluon, chiral,..condensates) 
indicates that a non-vanishing expectation value of the
Higgs field may correspond to a real medium
made up by the physical Bose condensation process of 
elementary spinless quanta whose `empty' vacuum state
is not the true ground state of the theory \cite{mech}. 
Noticing that bodies can flow without any apparent
friction in such a medium, it is natural to 
represent the Higgs condensate as a superfluid. 
In this perspective, such a physical vacuum should support long-wavelength
density fluctuations. In fact, the existence of density fluctuations
in any known medium
is a very general experimental fact, depending on the coherent response of
the elementary constituents to disturbances whose wavelength is much larger
than their mean free path \cite{seminar}. 
This leads to an universal description, the
`hydrodynamical regime', that does not depend on the details of the
underlying molecular dynamics. By accepting this argument, and quite 
independently of the Goldstone phenomenon, the energy spectrum 
of a Higgs condensate should terminate with 
an `acoustic' branch, say $\tilde{E}({\bf{p}})=c_s|{\bf{p}}|$
for ${\bf{p}} \to 0$, as for the propagation of sound waves in ordinary media. 

However, leaving aside the Goldstone bosons, i.e. for a spontaneously broken 
one-component $\lambda\Phi^4$ theory, the particle content of the broken phase
is usually represented as a single massive field, the (singlet) Higgs boson. 
Although there is no rigorous proof \cite{book}, 
the Fourier transform of the connected Euclidean propagator is assumed to
tend to a finite limit, say
$G(p) \to {{{\rm 1} }\over{M^2_h}}$ when the 4-momentum
$p \to 0$, and the mass squared $M^2_h$ 
is related to the quadratic shape of a semi-classical 
effective potential $V_{\rm NC}(\phi)$ (`NC'=non-convex)
at its
non-trivial absolute minima, say $\phi=\pm v$. Equivalently, 
the energy spectrum of the broken phase should
tend to a non-zero value, $\tilde{E}({\bf{p}}) \to M_h$, 
when ${\bf{p}} \to 0$ so that the non-zero quantity $\tilde{E}(0)=M_h$
gives rise to an exponential decay $\sim e^{-\tilde{E}(0)T}$ of the 
connected Euclidean propagator.

Clearly, by considering the broken phase as a `condensate', 
the idea of an energy spectrum $\sqrt{ {\bf{p}}^2 + M^2_h}$ down to 
${\bf{p}}=0$ seems unnatural. In fact, for very long wavelengths, one
would expect the lowest excitations to arise from small displacements of
the condensed quanta that already `pre-exist' in the ground state. 
Our idea of density fluctuations  is motivated by general considerations
that should be relevant in any medium and, in particular, in a
Bose system at zero temperature. To this end, and 
for the convenience
of the reader, we shall report the following quotations:

~~~ i)  "Any quantum liquid consisting of particles with integral 
spin (such as the liquid isotope $^4$He) must certainly have a spectrum of
this type...In a quantum Bose liquid, elementary excitations with small 
momenta ${\bf{p}}$  (wavelengths large compared with distances between atoms) 
correspond to ordinary hydrodynamic sound waves, i.e. are phonons. This 
means that the energy of such quasi-particles is a linear function of their
momentum" \cite{pita}.
 
~~~ii)  " We now come to the key argument of superfluidity: the only 
low-energy excitations are phonons. Phonons are excited states of compression,
or states involving small displacements of each atom with a resultant
change in density" \cite{fey}.

~~~iii) "We have seen that low-energy non-phonon excitations are 
impossible. In other words, there are no possible long-distance movements of
the atoms that do not change the density" \cite{fey}.

After this preliminary introduction, 
we shall point out that the
apparent contradiction between the conventional picture of symmetry breaking
and the physical expectation of 
a superfluid medium with density fluctuations has a precise counterpart in 
a basic non-perturbative 
ambiguity for the $p \to 0$ limit of the inverse connected propagator
$G^{-1}(p)$. This is a two-valued function when $p \to 0$ and
includes the case $G^{-1}(p=0)=0$, as in a gap-less theory. 
This ambiguity, by itself, 
 does not {\it prove} that the energy spectrum is actually $\sim c_s|{\bf{p}}|$ 
for ${\bf{p}} \to 0$, nor it provides the value of $c_s$. 
 However, it represents a purely quantum-field
theoretical argument in favour of the existence of a gap-less branch and of
the intuitive picture of the broken phase
as a real physical medium with density fluctuations. 

Before reporting any calculation, let us first try to understand why 
the vacuum of a `pro forma' Lorentz-invariant quantum field theory may be
such kind of medium. This question may have several answers.
For instance, a fundamental phenomenon as
the macroscopic occupation of the same quantum state 
(say ${\bf{p}}=0$ in some frame) may 
represent the operative construction of 
a `quantum aether' \cite{dirac,volovik}. This would be
quite distinct from the 
aether of classical physics whose constituents were
assumed to follow definite space-time trajectories.
However, it would also be different
from the empty space-time of special relativity, assumed at the base of
axiomatic quantum field theory to deduce the exact Lorentz-covariance of
the energy spectrum.

In addition, one should take into account the 
approximate nature of locality in
cutoff-dependent quantum field theories. 
In this picture, the elementary quanta are treated as `hard spheres', 
as for the molecules of ordinary matter. Thus, the notion of the
vacuum as a `condensate' acquires an intuitive physical meaning. 
For the same reason, however, the simple idea that deviations from 
Lorentz-covariance take only place at the cutoff scale may be incorrect:
non-perturbative vacuum condensation may
give rise to a hierarchy of scales
such that the region of Lorentz-covariance is sandwiched {\it both}
by the high- and low-energy regions.

In fact, in general, 
an ultraviolet cutoff induces vacuum-dependent
{\it reentrant violations of
special relativity in the low-energy corner} \cite{volo2}. In the simplest
possible case, these extend over a small shell of momenta, say 
$|{\bf{p}}|< \delta$, where the energy spectrum $\tilde{E}({\bf{p}})$ may
deviate from a Lorentz-covariant form and be distorted into a sound-wave
shape. However, 
Lorentz-covariance becomes an exact symmetry in the local limit. Therefore, 
for very large but finite $\Lambda$, one expects 
the scale $\delta$ to be naturally
infinitesimal in units of the energy scale associated
with the Lorentz-covariant part of the energy spectrum, say $M_h$. 
By introducing dimensionless quantities, the requirement of asymptotic 
Lorentz-covariance introduces 
a tight infrared-ultraviolet connection since
$\epsilon\equiv {{\delta}\over{M_h}} \to 0$ when
$t \equiv {{\Lambda}\over{M_h}} \to \infty$. In this sense, 
formally, 
${\cal O}({{\delta}\over{M_h}})$ vacuum-dependent corrections would be
equivalent to 
 ${\cal O}({{M_h}\over{\Lambda}})$ effects and these are always
neglected when discussing \cite{nielsen} how
Lorentz-covariance emerges at energy scales that are 
much smaller than the ultraviolet 
cutoff. Therefore, in the condensed phase of a cutoff theory, 
although Lorentz-covariance is formally recovered in the local
limit, one should expect infinitesimal deviations
in an infinitesimal region of momenta. 

In this context, one may ask what the word {\it infinitesimal} actually 
means in the physical world. For instance, by assuming
$\Lambda=10^{19}$ GeV and $M_h=250$ GeV, a scale 
$\delta=10^{-5}$ eV, for which
$\epsilon\equiv {{\delta}\over{M_h}}=4\cdot10^{-17} $, might well represent
the physical 
realization of a formally infinitesimal quantity. If this were
the right order of magnitude, 
the non-Lorentz-covariant density fluctuations of the vacuum would start
to show up 
from wavelengths larger than a centimeter up to infinity. These lengths
are actually {\it infinitely} large as compared 
 to the Fermi scale but have, nevertheless, a physical meaning. 
At the same time, the associated long-range interactions would have 
a strength $\epsilon^2 \sim {{M^2_h}\over{\Lambda^2}} =
{\cal O}(10^{-33}) $ relatively to the Fermi constant. Although small, 
this strength is non-vanishing and these interactions 
can play a physical role on macroscopic distances.

We shall now follow, in Sects. 2 and 3, 
two different methods to display the ambiguity in the zero-momentum limit
of the connected propagator in the broken phase.
In Sect.4 we shall present our conclusions and a brief discussion of the
most general consequences of our results.

\section{ The functional integration over the background field}

When discussing spontaneous symmetry breaking, the starting point 
is the separation of the scalar field into a constant background
and a shifted fluctuation field, namely 
\BE
\label{shift}
          \Phi(x)= \phi + h(x)
\EE
In order Eq.(\ref{shift}) to be unambiguous, 
$\phi$ denotes the spatial average in a large 4-volume $\Omega$ 
\BE
\label{average}
          \phi= {{1}\over{\Omega}}\int d^4x~ \Phi(x)
\EE
and the limit $\Omega \to \infty$ has to be taken at the end.

In this way, the full functional measure can be expressed as
\BE
\label{measure}
                \int[d\Phi(x)]...=\int^{+\infty}_{-\infty}d\phi\int[dh(x)]...
\EE
and the functional integration on the r.h.s. of Eq.(\ref{measure}) 
is over all quantum modes with 4-momentum $p\neq 0$. 

After integrating 
out all non-zero quantum modes, 
the generating functional in the presence of a space-time constant 
source $J$ is given by
\BE
\label{zetaj}
      Z(J)= \int^{+ \infty}_{-\infty} d\phi~ \exp [-\Omega
(V_{\rm NC}(\phi) - J\phi)]
\EE 
and
$V_{\rm NC}(\phi)$ denotes the usual non-convex (`NC') effective potential
obtained order by order 
in the loop expansion. Finally, by introducing the generating functional for
connected Green's functions $w(J)$ through
\BE
\label{log}
\Omega~ w(J)=\ln {{Z(J)}\over{Z(0)}}
\EE
one can compute the field expectation value
\BE
\label{phij}
\varphi(J)={{dw}\over{dJ}}
\EE
and the zero-momentum propagator
\BE
\label{GJ0}
G_J(p=0)={{d^2w}\over{dJ^2}}
\EE
In this framework, 
spontaneous symmetry breaking corresponds to non-zero values of 
Eq.(\ref{phij}) in the double limit $J \to \pm 0$ and 
$\Omega \to \infty$. 

Now, by denoting $\pm v$ the absolute minima of 
$V_{\rm NC}$ and 
$M^2_h=V''_{\rm NC}$ its quadratic shape at these extrema, one usually assumes 
\BE
\label{ssb}
\lim_{\Omega \to \infty} \lim_{J \to \pm 0} \varphi(J) = \pm v
\EE
{\it and}
\BE
\label{GA}
\lim_{\Omega \to \infty} \lim_{J \to \pm 0} G_J(p=0)={{1}\over{M^2_h}}
\EE
In this case, the excitations in the broken phase 
would be massive particles (the conventional Higgs bosons)
whose mass $M_h$ is determined by the positive 
curvature of $V_{\rm NC}$ at its absolute minima. However, 
at $\varphi=\pm v$, 
besides the value ${{1}\over{M^2_h}}$, one also finds \cite{legendre}
\BE
\label{GB}
\lim_{\Omega \to \infty} \lim_{J \to \pm 0} G_J(p=0)=+\infty
\EE
a result that has no counterpart in perturbation theory. 

Let us review how this result emerges from
the saddle-point approximation, valid for $\Omega \to \infty$. In this case,
we get
\BE
\label{saddle}
w(J)= {{J^2}\over{2 M^2_h}} + 
{{\ln \cosh(\Omega J v)}\over{\Omega}}
\EE
and
\BE
\label{phij2}
\varphi={{dw}\over{dJ}}= {{J}\over{M^2_h}} + v \tanh(\Omega J v)
\EE
\BE
\label{GJ}
G_J(p=0)={{d^2w}\over{dJ^2}}=
{{1}\over{M^2_h}} + {{\Omega v^2}\over{\cosh^2(\Omega Jv)}}
\EE
To determine the zero-momentum propagator in a given background $\varphi$, 
we should now invert $J$ as a function of $\varphi$ from Eq.(\ref{phij2})
and replace it in (\ref{GJ}). However, 
being interested in the limit $J \to 0$ it is easier to look for the
possible limiting behaviours of (\ref{GJ}). 

Since both $J$ and $\Omega$ are dimensionful quantities, it is convenient 
to introduce dimensionless variables
\BE
\label{exj}
        x\equiv \Omega J v
\EE    
and
\BE
\label{uai}
          y\equiv \Omega v^2 M^2_h
\EE
so that Eqs.(\ref{phij2}) and (\ref{GJ}) become
\BE
\label{phij3}
   \varphi = v [ {{x}\over{y}} + \tanh(x)]
\EE
and
\BE
\label{GJ2}
G_J(p=0) = {{1}\over{M^2_h}}[1 + {{y}\over{\cosh^2(x)}}]
\EE
In this representation, taking the two limits 
$\Omega \to \infty$ and $J \to \pm 0$ correspond to choose some path in the 
two-dimensional space $(x,y)$. The former gives trivially $y \to \infty$.
The latter, on the other hand, is equivalent to $ {{x}\over{y}} \to \pm 0$ since
\BE
{{J}\over{ v M^2_h}} = {{x}\over{y}} 
\EE
with many alternative possibilities. If we require a non-zero 
limit of $\varphi$ this amounts to an asymptotic non-zero value of $x$. If this 
value is finite, say $x=x_o$ we get asymptotically
\BE
\label{finitephi}
    \varphi \to v \tanh(x_o)
\EE
and
\BE
\label{finiteg}
G_J(p=0) \to {{1}\over{M^2_h}} {{y}\over{\cosh^2(x_o)}} \to \infty
\EE
implying the existence of gap-less modes for every non-zero value of $\varphi$. 
On the other hand, if $ x \to \pm \infty$ we obtain
\BE
\varphi \to \pm v
\EE
In this case $G_J(p=0)$ tend to ${{1}\over{M^2_h}}$ 
(to $+\infty$) depending on whether $y$
diverges slower (faster) than $\cosh^2(x)$. 

The above results admit a simple
geometrical interpretation in terms of the shape of the
effective potential $V_{\rm LT}(\varphi)$ as defined from the Legendre 
transform (`LT') of $w(J)$.
After obtaining $J$ as a function of
$\varphi$ from Eq.(\ref{phij2}), the inverse zero-momentum propagator
in a given background $\varphi$ is related to the second-derivative of the 
Legendre-transformed effective potential, namely
\BE
\label{step1}
G^{-1}_\varphi(p=0)= {{dJ}\over{d\varphi}}=
       \frac{ d^2 V_{\rm LT}}{d \varphi^2} 
\EE
In this case, Eqs.(\ref{finitephi}) and (\ref{finiteg}) require
a vanishing result from Eq.(\ref{step1}) when $ -v < \varphi < v$. This 
is precisely what happens since $V_{\rm LT}$ 
becomes flat in the region enclosed by the absolute minima of the non-convex
effective potential when
$\Omega \to \infty$. This is the usual `Maxwell construction' where
$V_{\rm LT}(\varphi)=V_{\rm NC}(\pm v)$, 
for $-v \leq \varphi \leq v$, and 
$ V_{\rm LT}(\varphi)=V_{\rm NC}(\varphi)$ for $\varphi^2 > v^2$.

Notice, however, that
the limit of (\ref{step1}) for $\varphi \to \pm v$
cannot so simply be identified with $M^2_h$. In fact, even within 
the `Maxwell construction', this identification requires
a strong additional assumption: 
the derivative in Eq.(\ref{step1}) has to be a left- (or right-)
derivative depending on whether we consider the point $\varphi=-v$ 
( or $\varphi=+ v$). Now, this is just a prescription since
derivatives depend on the chosen path (unless one deals with infinitely 
differentiable functions) and, differently from $V_{\rm NC}$, the
Legendre transformed $V_{\rm LT}$ is not an infinitely differentiable
function in the presence of spontaneous symmetry breaking \cite{syma}.

Therefore, in general, 
Eq.(\ref{step1}) leads to multiple
solutions at $\varphi=\pm v$. Namely, an exterior derivative for which
$G^{-1}_{\rm ext}(0)=M^2_h$ but also a
$G^{-1}_{\rm int}(0)=0$, as when approaching the points
$\pm v$ from the internal region 
where the Legendre-transformed potential becomes flat for $\Omega \to \infty$.
 These two different alternatives correspond to the various 
limits $y \to \infty$ and $x \to \pm \infty$ in Eq.(\ref{GJ2}) such that
${{y}\over{\cosh^2(x)}}$ tends to zero or infinity.

We observe that the `Maxwell construction', i.e. the
replacement $V_{\rm NC} \to V_{\rm LT}$ as a genuine quantum effect, 
was also discovered in ref.\cite{branchina}. Graphically,
the resulting effective potential becomes flatter and flatter between $-v$ and
$+v$ when removing the infrared cutoff. Numerically, 
the ratio between left- and right- second
derivatives at the absolute minima of $V_{\rm NC}$ is found to
diverge in the same limit \cite{private}.

We conclude this section with the remark that the singular zero-momentum
behaviour we have pointed out does not depend at any stage on the existence
of a continuous symmetry of the classical potential. 
As such, there are no differences in a spontaneously
broken O(N) theory. Beyond the approximation where
the `Higgs condensate' is treated as a purely
classical background, one has to 
perform one more integration over the zero-momentum mode of the condensed
$\sigma-$field. Therefore, 
all ambiguities in computing the inverse propagator of the
$\sigma-$field through Eq.(\ref{step1}) remain. In this sense, the
possibility of multiple values for $G_\sigma(p=0)$ has 
nothing to do with the number of field components. 

\section{ Re-summation of the tadpole graphs}

The possibility of a divergent zero-4-momentum propagator in the broken
phase, as
illustrated in the previous section, is non-perturbative and
independent of any diagrammatic analysis. As an additional
evidence for the subtle nature of the $p \to 0$ limit of $G(p)$, 
we shall attempt, however, to isolate the possible
origin of this effect in the one-particle reducible
zero-momentum tadpole graphs. These enter the 
usual diagrammatic expansion in the presence of
a constant background field and can be considered a manifestation of
the quantum nature of the scalar condensate. 
The expansion we shall consider 
is defined in terms of the 1PI graphs generated by
the non-convex 
effective potential $V_{\rm NC}(\phi)$ considered before. In this respect,
the tadpole graphs are fully 
non-perturbative and have to be re-summed 
to all orders. The genuine 1PI interaction graphs, on the other hand, 
represent perturbative effects and
can be considered to any desired order in the loop expansion, without 
changing qualitatively the conclusions. In the following, after some
preliminaries, we shall address the zero-momentum 
propagator at the absolute minima of
$V_{\rm NC}(\phi)$.

Let us start by defining 
\BE
\label{vefft}
{{dV_{\rm NC}}\over{d \phi}} \equiv J(\phi)\equiv \phi T(\phi^2)
\EE
and $\phi=\pm v \neq 0$ are the solutions of
\BE
\label{1point}         
T(\phi^2)=0
\EE
with
\BE
\label{second}
   \left. \frac{ d^2 V_{\rm NC}}{d \phi^2}\right|_{\phi=\pm v} >0
\EE
Usually, one defines the $h-$field propagator from a Dyson sum of
1PI graphs only, say
\BE
\label{1PI}
G(p)|_{\rm 1PI} \equiv D(p)
\EE
where ($D^{-1}(0) \equiv D^{-1}(p=0))$
\BE
\label{ident}
D^{-1}(0) \equiv
   {{ d^2 V_{\rm NC} }\over{d \phi^2}}
\EE 
This 
provides the conventional definition of $M^2_h$ through 
Eq.(\ref{ident}) at $\phi =\pm v$. 

In this description one
neglects the possible role of the
one-particle reducible, zero-momentum tadpole graphs. The reason is  that 
their sum is proportional to the 1-point function, i.e. to 
$J(\phi)$ in Eq.(\ref{vefft}) 
that vanishes by definition at $\phi=\pm v$. 
However, the zero-momentum tadpole subgraphs are attached to the other parts
of the  diagrams through zero-momentum propagators so that, in an all-order 
calculation, their overall contribution vanishes 
provided the full propagator $G(0)$ is non-singular at the 
minima. In this respect, neglecting the
tadpole graphs amounts
to {\it assume} the regularity of 
$G(0)$ at $\phi=\pm v$ which is certainly
true in a finite-order expansion. In an intuitive analogy, when
$\phi\to \pm v$, $J(\phi)$ represents an infinitesimal driving force due to
the medium. Thus, it will not produce any observable effect, 
unless the mass of a body vanishes in the same limit. The 
complication in our case is that 
the mass of our `body', the inverse propagator $G^{-1}(0)$, 
 depends on the medium and on the driving force 
itself. 

For this reason, we shall try 
to control the full propagator in a small region of 
$\phi-$ values around the minima, by
including all zero-momentum 
tadpole graphs, and finally take the limit $\phi \to \pm v$.
We observe that the problem of tadpole graphs 
was considered in ref.\cite{munster} where the emphasis was mainly 
to find an efficient way to re-arrange the perturbative expansion.  Here,
we shall attempt a non-perturbative all-order re-summation of the various 
effects to check the regularity of $G(0)$ for $\phi \to \pm v$.

We shall approach the problem in two steps.
In a first step, we shall consider the contributions to the propagator
by including
 all possible insertions of zero-momentum lines in the internal part of the
graphs, i.e. inside 1PI vertices. 
At this stage, however, 
the external zero-momentum propagators to the sources 
maintain their starting value $D(0)$ {\it at} $J=0$.
This approximation gives rise to an auxiliary inverse propagator 
given by 
\BE
\label{aux1}
{G}_{\rm aux}^{-1}(p)= 
D^{-1}(p)
-\varphi z \Gamma_3(p,0,-p)
+{{(\varphi z)^2}\over{2!}}\Gamma_4(0,0,p,-p)
-{{(\varphi z)^3}\over{3!}}\Gamma_5(0,0,0,p,-p)+..
\EE
where
\BE
\label{basic}
             z\equiv T(\phi^2) D(0)
\EE
represents the basic one-tadpole insertion.
Eq.(\ref{aux1}) can be easily
checked diagrammatically starting from
the tree approximation where 
\BE
\label{tree}
V_{\rm tree}={{1}\over{2}} r \phi^2 + {{\lambda}\over{4!}} \phi^4 
\EE
\BE
D^{-1}(p)= p^2 + r + {{\lambda\phi^2}\over{2}} 
\EE
and
$\Gamma_3(p,0,-p)= \lambda\phi$, 
$\Gamma_4(0,0,p,-p)= \lambda$
(with all $\Gamma_n$ vanishing for $n>4$). 

Now, by using the relation of the zero-momentum 1PI
vertices with the effective potential at an arbitrary $\phi$
\BE
       \Gamma_n(0,0,...0)=
        { { d^n V_{\rm NC} }\over {d \phi^n}}
\EE
we can express the auxiliary
zero-4-momentum inverse propagator of Eq.(\ref{aux1}) as
\BE
\label{aux2}
G_{\rm aux}^{-1}(0)= 
       \left. \frac{ d^2 V_{\rm NC}}{d \phi^2} 
\right|_{ {\phi_{\rm aux}}=\phi(1 - z) } 
\EE
For $\phi \to \pm v$, this partial re-summation of tadpole graphs
gives a vanishing contribution to the inverse propagator so that
$G_{\rm aux}^{-1}(0) \to M^2_h$ as determined from the quadratic
shape of $V_{\rm NC}$ at its absolute minima. However, for arbitrary $\phi$
even this partial re-summation produces non-perturbative modifications of the
zero-momentum propagator. For instance, as one can check with the 
tree-level potential, there are values of 
$\phi$ where $D^{-1}(0)$ is positive but 
$G_{\rm aux}^{-1}(0)$ is negative.

The second step 
consists in including now
all possible tadpole corrections on each external zero-momentum line.
In fact, in a diagrammatic expansion,  a single external zero-momentum
leg gives rise to a new
 infinite hierarchy of graphs, each producing
another infinite number of graphs and so on. 
Despite of the apparent complexity of the task, 
the final outcome of this computation can
be cast in a rather simple form, at least on a formal ground. The point is
that this infinite class of graphs can be included 
into a re-definition of the two basic expansion parameters entering the
tadpole re-summation: the source function $J$ and the zero-momentum 
propagator $D(0)$. 

In fact, to any finite order in $J$, 
we can re-arrange the expansion for the zero-momentum
propagator (all $\Gamma_n$ are evaluated at zero external momenta)
\BE
\label{infinite}
G(0)= D(0) + J\Gamma_3 D^3(0)+ {{3J^2\Gamma^2_3}\over{2}}D^5(0)-
{{\Gamma_4J^2}\over{2}}D^4(0) + {\cal O}(J^3)
\EE
in terms of a modified source
\BE
\label{modsource}
\tilde{J}= J - {{J^2\Gamma_3}\over{2}}D^2(0) +{\cal O}(J^3) \equiv 
\phi \tilde {T}(\phi^2)
\EE
in such a way that the formal power series for the
exact inverse zero-momentum propagator can be expressed as
\BE
\label{formal}
G^{-1}(0)= 
       \left. \frac{ d^2 V_{\rm NC}}{d \phi^2} 
\right|_{ {\hat{\phi}}=\phi(1 - \tau) } 
\EE
with
\BE
\label{tauu}
            \tau\equiv \tilde{T}(\phi^2)G(0)
\EE
i.e., as in (\ref{aux2}) with the replacement $z \to \tau$. 
As a result, after including tadpole graphs to all orders, 
one finds multiple solutions for the zero-4-momentum propagator 
that differ from (\ref{ident}) even when $\phi \to \pm v$.

The situation is similar to solving
the following equation
\BE
         f^{-1}(x)= 1 +x^2 - g^2 x^2 f(x)
\EE
For  $x \to 0$ there are two distinct limiting
behaviours : a) $ f(x) \to 1$ and b) $f(x) \sim {{1}\over{g^2x^2}} \to +\infty$.
However, only the former solution is recovered in a finite number of iterations
from
\BE
f_0(x)= {{1}\over{1+x^2}}
\EE
for $g^2=0$. 
In the case of $\lambda\Phi^4$ theory, deriving the gap-less mode 
from tadpole re-summation
corresponds to the b) type of behaviour.

In this sense, the re-summation in Eq.(\ref{formal})
is only formal since, with this procedure, the function 
$\tilde{J}(\phi)$ is always 
determined as a finite-order polynomial up to higher orders in $J$, 
as it happens in perturbation theory. However, this is not so important 
since the possibility of
a singular zero-4-momentum propagator at $\phi=\pm v$
does not depend on the {\it form} of 
$\tilde{J}(\phi)$ but only on its {\it vanishing} 
at $\phi=\pm v$. As we shall see, this vanishing reflects
simple geometrical properties of the non-convex effective potential 
$V_{\rm NC}$.

To understand this point, we first observe that traditionally, 
for $\phi \to \pm v$, 
tadpole re-summation has been considered to be irrelevant. Namely, in the
literature the 
inverse propagator is always defined from
Eq.(\ref{ident}) that neglects the tadpole graphs altogether. As anticipated
in the Introduction, this is the main motivation
to relate the physical Higgs mass to the quadratic shape of $V_{\rm NC}$.
Therefore, after including the tadpole graphs, 
Eq.(\ref{ident}) should also be contained in
Eq.(\ref{formal}), at least as a particular solution where 
$G(0)=D(0)$. To this end, however, when $ J \to 0$ 
and $\phi \to \pm v$, also 
the modified source $\tilde{J}$ has to vanish. 
The alternative possibility, i.e. that the full 
$\tilde{J}$ remains non-zero when $\phi \to \pm v$, would produce 
a drastic result. In fact, after tadpole re-summation, an inverse propagator
as in Eq.(\ref{ident}) would never be recovered from (\ref{formal}), 
even as a particular solution. This would also contradict the indications
of the previous section where we have found evidence for {\it both}
regular and singular values of the zero-4-momentum propagator at $\phi=\pm v$.

Adopting the natural point of view that 
the modified source $\tilde{J}$ (and so $\tilde{T}$) vanishes when
$\phi \to \pm v$, 
Eq.(\ref{formal}) provides a regular solution
${G}^{-1}_{\rm reg}(0)=D^{-1}(0)$ for which
\BE
\label{taureg}
           \lim_{\phi \to \pm v} \tau = \bar{\tau}=0
\EE
and a singular solution
${G}^{-1}_{\rm sing}(0)=0$ such that 
\BE
\label{tausing}
           \lim_{\phi \to \pm v} \tau= \bar{\tau} \neq 0
\EE
As an example, let us consider the situation of
$V_{\rm NC} \equiv V_{\rm tree}$ in Eq.(\ref{tree}) which is
equivalent to re-summing tree-level 
tadpole graphs to all orders (i.e. no loop diagrams).
In this case the regular solution is
${G}^{-1}_{\rm reg}(0)={{\lambda v^2}\over{3}}$, while the
singular solution is
\BE
\label{singular}
\lim_{\phi^2 \to v^2}
{G}^{-1}_{\rm sing}(0)=
{{\lambda v^2}\over{2}}
[ \bar{\tau}^2- 2\bar{\tau} + {{2}\over{3}}]=0
\EE
which implies limiting values 
$ \bar{\tau}=1 \pm {{1}\over{\sqrt{3}}}$.

In general, beyond the tree-approximation, 
finding the singular solution 
$G^{-1}_{\rm sing}(0)= 0$ at $\phi=\pm v$ is equivalent to determine that value of 
$\hat{\phi}^2\equiv v^2(1-\bar{\tau})^2$ where
${{d^2V_{\rm NC}}\over{d \phi^2}}=0$.
 For instance, in the case of the
Coleman-Weinberg effective potential
\BE
            V_{\rm NC}(\phi)= {{\lambda^2\phi^4}\over{256\pi^2}}
(\ln {{\phi^2}\over{v^2}} -{{1}\over{2}})
\EE
the required values are $\bar{\tau}=1 \pm e^{-1/3}$. In principle, 
such solutions exist in any
approximation to $V_{\rm NC}$ due to the
very general properties of the shape of a non-convex effective potential.

\section{ Conclusions and outlook}

In principle, a medium can support different types of excitations. For instance,
the energy spectrum of superfluid $^4$He is considered to arise from
the combined effect of two types of
excitations, phonons and rotons, 
whose separate energy spectra `match', giving rise to a complicate
pattern \cite{griffin}. In this sense, there is a unique spectrum but
phonons and rotons can be considered
different `particles' reflecting 
different aspects of the superfluid helium wave function \cite{fey}.
 
The essential point is that, for superfluid $^4$He, 
the existence of two types of
excitations  was first deduced theoretically by Landau
on the base of very general arguments \cite{landau}. According to this original
idea, there are phonons with energy
$E_{\rm ph}({\bf{p}}) = v_s |{\bf{p}}|$ and
rotons with energy
$E_{\rm rot}({\bf{p}})= \Delta + {{ {\bf{p}}^2 }\over{2\mu}}$. 
Only {\it later},
it was experimentally discovered that
there is a single energy spectrum $E({\bf{p}})$. This is made up
by a continuous matching of these two
different parts and is dominated by phonons for ${\bf{p}} \to 0$ where 
the rotons become unphysical.

Let us now consider our analysis of the zero-momentum 
propagator. Its two-valued nature may be the
indication that something similar happens in the broken phase, 
in agreement with the intuitive picture of
the Higgs vacuum as a superfluid system. In fact, 
the existence of both a $G^{-1}_a(0)\equiv M^2_h$ and a
$G^{-1}_b(0)\equiv 0$ implies that there would be
two possible types of excitations with the same quantum numbers but
different energies when the 3-momentum
${\bf{p}} \to 0$: 
a massive one, with $\tilde{E}_a({\bf{p}}) \to M_h$, and a gap-less one with 
$\tilde{E}_b({\bf{p}}) \to 0$ that, `a priori', can both propagate 
(and interfere) in the broken-symmetry phase. In analogy with $^4$He, we 
would conclude that the latter dominates the exponential decay 
$\sim e^{-\tilde{E}_b({\bf{p}})T}$  of the
connected euclidean correlator for
${\bf{p}} \to 0$ so that 
the massive excitation becomes unphysical in the infrared region.
Therefore, differently from the simplest perturbative indications, 
in a (one-component) spontaneously broken
$\lambda\Phi^4$ theory there would be no energy-gap associated
with the `Higgs mass'  $M_h$, as in a genuine massive single-particle 
theory where the relation
\BE
\label{Ea}
\tilde{E}_a({\bf{p}})=\sqrt{ {\bf{p}}^2 + M^2_h} 
\EE
remains true for ${\bf{p}} \to 0$. Rather, the far-infrared region would
be dominated by gap-less collective excitations whose typical energy spectrum
for  ${\bf{p}} \to 0$ 
\BE
\label{Eb}
\tilde{E}_b({\bf{p}}) \equiv c_s |{\bf{p}}| 
\EE
depends on an unknown parameter $c_s$. This, 
according to the arguments given in the Introduction,
would represent the `sound velocity' 
for the density fluctuations of the superfluid scalar condensate. 

By their very nature, these density fluctuations
represent non-Lorentz covariant effects and, following the discussion
given in the Introduction, should be restricted to  
an infinitesimal region 
of momenta $|{\bf{p}}| < \delta$ with 
${{\delta}\over{M_h}}={\cal O}({{M_h}\over{\Lambda}})$, $\Lambda$ being 
the ultraviolet cutoff of the theory. 

As discussed in the
Introduction, the strength of the associated
long-range interactions is also expected to be infinitesimal. In this
framework, 
one may consider possible viable phenomenological frameworks, of the type
presented in ref.\cite{giudice2}, where the massive branch 
dominates at higher momenta as
it would happen in a superfluid system where $\Delta=\mu\equiv M_h$.
The superfluid analogy is further supported by the observation
\cite{mech} that, as for the 
interatomic $^4$He-$^4$He potential, the low-energy limit of cutoff
$\lambda\Phi^4$ is 
a theory of quanta with a short-range repulsive core and
a long-range attractive tail. 
The latter originates from ultraviolet-finite parts
of higher loop graphs \cite{mech} that give rise to a 
$- {{\lambda^2 e^{-2mr} }\over{r^3}}$ attraction, $m$ being the mass of the elementary
condensing quanta. Differently from the usual ultraviolet divergences, 
this finite part cannot be reabsorbed into a standard 
re-definition of the tree-level, repulsive $+\lambda\dl$ contact 
potential and is essential for a physical description of the
condensation process 
when approaching the phase transition limit $m \to 0$ where the symmetric 
vacuum at $\langle\Phi\rangle=0$ becomes unstable. 

Of course, one may object that we have not provided
a calculation of the energy spectrum. 
Admittedly, this is a weak point of
our analysis that just concentrates on the zero-momentum propagator. 
 At the same time, it will not be so easy to improve on it.
In fact, for a full calculation of the energy spectrum, 
 one should improve on 
the usual covariant generalization of the Bogolubov method used in
ref.\cite{mech}.
This approximation, where the creation and annihilation operators 
$a_{ {\bf{p}}=0}$, $a^{\dagger}_{ {\bf{p}}=0}$
for the elementary quanta in the
${\bf{p}}=0$ mode are simply replaced by the c-number $\sqrt{N}$, is
equivalent to treat the scalar condensate as a purely classical background
field $\phi$ entering the 
quadratic hamiltonian for the shifted fluctuation field $h(x)$. In this case,
the energy spectrum is just $\sqrt{ {\bf{p}}^2 + M^2_h}$ and, therefore, 
the existence of a gap-less excitation
branch for ${\bf{p}} \to 0$ could not be discovered there.

However, in ref.\cite{mech} it was noticed that the Bogolubov method 
does not allow for a straightforward extrapolation 
down to ${\bf{p}}= 0$.  This is due to an ambiguity relating 
the original creation and annihilation 
operators to their Bogolubov-transformed counterpart in the limit 
${\bf{p}} \to 0$. Requiring continuity
of the massive energy spectrum down to ${\bf{p}}=0$, 
is equivalent to replace
$a_{ {\bf{p}}=0}$ and
$a^{\dagger}_{ {\bf{p}}=0}$
with $\sqrt{N}$. This choice is the second-quantization
analog of `freezing' $\phi=\pm v$ without performing the functional integration
in Sect. 2 or without first re-summing the zero-momentum tadpole 
graphs in Sect. 3. In this case, the only 
solution is $G^{-1}(0)=M^2_h$. 

In this sense, one should  first
include a genuine operator 
part for $a_{ {\bf{p}}=0}$ \cite{ezawa}, say
$a_{ {\bf{p}}=0 }=\sqrt{N}-\hat{\xi}$ with 
$\langle\hat{\xi}\rangle= \alpha$ and $|\alpha|^2\ll N$.
This introduces new contributions as the
3-linear couplings
$a^{\dagger}_{ {\bf{p}}} a_{ {\bf{p}}} \hat{\xi} $,...the 4-linear
couplings
$a^{\dagger}_{ {\bf{p}}} a_{ {\bf{p}}} \hat{\xi}^\dagger \hat{\xi} $,...
whose effects should be preliminarly computed
in analogy with the zero-momentum tadpole graphs 
considered in Sect.3. 
The new contributions produce corrections to 
the standard Bogolubov massive spectrum computed 
in ref.\cite{mech}.

Now, the stationarity value of $\alpha$, say 
$\alpha=\bar{\alpha}$, has to be
determined after minimizing the energy density and does not necessarily 
correspond to
$\alpha= 0$. This can easily be understood, by noticing that, after
minimization, the number of particles in the condensate is 
\BE
\langle a^{\dagger}_{ {\bf{p}}=0 }a_{ {\bf{p}}=0 } \rangle= N -
2 Re(\bar{\alpha}) \sqrt{N} + {\cal O}(\bar{\alpha}^2)
\EE
Therefore, $\alpha$ describes dynamical re-arrangements of
the total number of particles between the condensate and the states with
${\bf{p}}\neq 0$. By {\it dynamical}, we mean that a non-zero $\alpha$ 
changes the fraction of particles in the condensate and that this modification
is not an overall change of $N$. The latter
does not depend on the occurrence of Bose condensation but concerns the
infinite volume limit of any system with a given particle density. 

With an intuitive term, a
non-zero $\bar{\alpha}$ produces a {\it depletion} of the condensate,
i.e. additional
contributions to the ground state wave function as
$({\bf{p}}, -{\bf{p}})$, $({\bf{p}}_1,{\bf{p}}_2,-{\bf{p}}_1-{\bf{p}}_2)$,...
These are needed for the dynamical equilibrium in the presence of interactions
and 
extend over a typical region of momenta, say
$|{\bf{p}}| < \delta$, such that
$\epsilon\equiv {{\delta}\over{M_h}} \to 0 $ when 
$\bar{\alpha} \to 0$. In this limit, in fact, the 
Bogolubov spectrum $\sqrt{ {\bf{p}}^2 + M^2_h}$ applies
to the whole range of momenta producing an exact Lorentz-covariant
theory (with the possible exception of the zero-measure 
set ${\bf{p}}=0$). 

Therefore, taking into account our discussion in the 
Introduction, we conclude that the limit of vanishing interactions, 
$\bar{\alpha} \to 0$, has to correspond to the continuum limit 
$t={{\Lambda}\over{M_h}} \to \infty$ of cutoff $\lambda\Phi^4$ theory, i.e.
`triviality', in full agreement with all rigorous results \cite{book}.
However, 
our analysis shows that,in the broken-symmetry phase, 
the approach to the continuum theory is
 more subtle than what is generally believed. In fact, usually, one just 
considers the deviations from `triviality' to be
perturbative corrections to a free massive theory, 
without any qualitative change for 
${\bf{p}} \to 0$.

We end up, by mentioning that the existence of gap-less modes in the broken 
symmetry phase finds some support in the results of numerical simulations.
These have been performed \cite{cea3} 
in the low-temperature phase of a one-component
4D Ising model. The lattice data for the exponential decay 
of the connected correlator show that, by simply increasing the lattice size,
one finds smaller and smaller values of the energy gap 
$\tilde{E}({\bf{p}}=0)$. Namely, by using 
the same lattice parameters that 
on a $20^4$ lattice give \cite{jansen}
$\tilde{E}(0)=0.3912 \pm 0.0012$, the results of ref.\cite{cea3} were
$\tilde{E}(0)=0.3826 \pm 0.0041$
on a $24^4$ lattice, 
$\tilde{E}(0)=0.3415\pm 0.0085$
on a $32^4$ lattice and
$\tilde{E}(0)=0.303\pm 0.026$
on a $40^4$ lattice. 

However, in re.\cite{cea3} the extraction of the energy-gap was obtained from
a fit to a single exponential 
$\sim e^{-\tilde{E}(0)T}$ for all time-slices. Although the quality of the 
fit was quite good, strictly speaking, 
due to the contamination from higher excited states, 
the energy-gap should
be obtained from the exponential decay for {\it asymptotic} $T$. 
In this case, by restricting to large $T$ only, 
a good signal-to-noise ratio requires 
a very high statistics (say 10 millions of sweeps) so that 
on large $32^4$ and $40^4$ lattices, one has to 
wait for very long running times.
For this reason, the numerical evidence
from lattice simulations, although promising, is still inconclusive and
suggest the need for additional efforts by other groups. 
\vskip 10 pt
{\bf Acknowledgements}~I thank J. Polonyi and P. M. Stevenson for useful
discussions.
\vfill
\eject

\end{document}